\documentstyle[preprint,aps]{revtex}

\begin{document}

\draft

\tighten

\preprint{
\begin{tabular}{l}
{\bf HIP-1997-62/TH} \\
{\tt hep-th/9607208}
\end{tabular}
}

\title{Two-loop Finiteness of Chern-Simons Theory in Background Field Method}

\author{M. Chaichian and W. F. Chen }

\address{ High Energy Physics Division, Department of Physics \\
 and\\ 
Helsinki Institute of Physics, P.O. Box 9 (Siltavuorenpenger 20 C)\\ 
FIN-00014, University of Helsinki, Finland}

\maketitle

\begin{abstract}
We perform two-loop calculation of Chern-Simons in background field method 
using the hybrid regularization of higher-covariant derivative and dimensional 
regularization. It is explicitly shown that Chern-Simons field theory is 
finite at the two-loop level. This finiteness plays an important role in 
the relation of Chern-Simons theory with two-dimensional conformal field
theory and the description of link invariant.  
\end{abstract}

\begin{flushleft}
{\bf 1. ~Introduction}
\end{flushleft}

The finiteness of $2+1$-dimensional Chern-Simons field theory (CS) has played 
a crucial role in its connection with 2-dimensional conformal invariant
field theory and in describing link invariant$\cite{witt,laba,gua}$. 
The  one-loop perturbative finiteness 
has been explicitly shown in various regularization schemes despite the fact
that there exists the dispute on the finite renormalization
of the coupling constant$\cite{alv}$. 
In fact, from the general analysis of the analytic property of the one-loop
amplitude by Speer$\cite{spe}$, the perturbative
finiteness of CS at one-loop is somehow trivial since it is
defined in odd dimensional space-time. Only when one considers the
quantum corrections at two-loop or higher-order, the UV divergences may appear.
Thus it is necessary to work in a concrete regularization scheme
and show whether the UV divergences really do or do not cancel. 
Although the UV finiteness of CS has been verified to 
every order in a regularization independent way$\cite{blasi}$, 
up to now there have appeared very few papers which 
deal with its two-loop or higher order perturbative behaviour.

In this letter, we shall explicitly show the two-loop perturbative 
finiteness of CS in the framework of background 
field method and by choosing the hybrid regularization
scheme of higher-covariant derivative 
and dimensional regularization. The advantage of adopting the background field
method is obvious: owing to explicit background gauge invariance$\cite{abb}$,
by considering only the two-point function of background
field, one can determine the local quantum effective action.
We shall choose the simplest higher-covariant derivative term, 
Yang-Mills action, which means that we start from the dimensional regulated
topologically massive Yang-Mills theory (TMYM).  From the viewpoint of
regularization,  the theory has two regulators, one is the dimensional
regulator, ${\epsilon}=3-n$, and the other is the topological 
mass $m$. Therefore,
the two-loop finiteness is equivalent to the existence of the limits 
${\epsilon}{\longrightarrow}0$ and $m{\longrightarrow}{\infty}$. 

In Sect. II we briefly review the definition of dimensional continuation,
which should receive a delicate consideration due to the 
antisymmetric tensor since it has been shown the naive dimensional 
regularization 
cannot make the theory well defined$\cite{bos}$. 
Then we give the Feynman rules
in the Landau background covariant gauge.  Sect. III contains the explicit
calculations and the analysis of two-loop two-point function of background
gauge field of topologically massive Yang-Mills theory. We show that
the $1/\epsilon$  terms cancel exactly.  The explicit background gauge 
invariance means that the three-point and other multi-point functions
should also contain no pole terms. Further we make use of
the cancellation theorem of IR divergence for Landau 
gauge given in Ref.$\cite{rao}$
to argue the existence of large topological mass limit and hence
the two-loop finiteness of CS is verified. 
Sect. IV presents the conclusion and some
discussions. 

\begin{flushleft}
{\bf 2. ~Feynman Rules of TMYM in Background Covariant Gauge}
\end{flushleft}

We start from the dimensionally regulated TMYM (in Euclidean space),
\begin{eqnarray}
S_{\rm m}[{\cal A}]=-i\mbox{sgn}(k){\int}d^3x{\epsilon}^{\mu\nu\rho}
\left[\frac{1}{2}{\cal A}_{\mu}^a{\partial}_{\nu}{\cal A}_{\rho}^a
+\frac{1}{3!}gf^{abc}{\cal A}_{\mu}^a{\cal A}_{\nu}^b{\cal A}_{\rho}^c\right]
+\frac{1}{4m}{\int}d^nx{\cal F}_{\mu\nu}^a{\cal F}^{{\mu\nu}a},
\end{eqnarray}
which is in fact the hybrid regularization of dimensional and
higher-covariant derivative regularization of CS theory.
The $n$-dimensional continuation of ${\epsilon}_{\mu\nu\rho}$ adopted here
is the one proposed by 't Hooft and Veltman$\cite{veltman2}$ and 
further by Breitenlohner and Maison$\cite{maison}$. For the Chern-Simons
 type theories, it was given explicitly in $\cite{mar1}$, 
\begin{eqnarray}
&&{\epsilon}_{\mu_1\mu_2\mu_3}{\epsilon}_{\nu_1\nu_2\nu_3}
=\sum_{\pi{\in}P_3}sgn(\pi){\Pi}_{i=1}^3\tilde{g}_{\mu_i\nu_{\pi(i)}},
\nonumber\\[2mm]
&&u_{\mu}=\tilde{u}_{\mu}{\oplus}\hat{u}_{\mu}, ~
g_{\mu\nu}=\tilde{g}_{\mu\nu}{\oplus}\hat{g}_{\mu\nu},~
\tilde{g}_{\mu\nu}\hat{u}^{\nu}=\hat{g}_{\mu\nu}\tilde{u}^{\nu}=0,
\nonumber\\[2mm]
&&\hat{g}^{\mu}_{~\mu}=n-3, ~\tilde{g}^{\mu}_{~\mu}=3, ~~~{\mu},
 {\nu}=1, 2,{\cdots}n,
\label{eq2}
\end{eqnarray}
where $g_{\mu\nu}={\delta}_{\mu\nu}$ is the Euclidean 
metric on $R^n$ and $\tilde{g}_{\mu\nu}$
and $\hat{g}_{\mu\nu}$ are its projections onto the subspaces $R^3$ and
 $R^{n-3}$. Note that this definition of dimensional continuation actually
makes the theory possess $SO(3){\otimes}SO(n-3)$ invariance rather than
$SO(n)$ $\cite{veltman2,mar1}$. To our knowledge, 
this prescription is the unique algebraically 
consistent definition for
dimensional continuation. In the calculations, the integer $n$
is promoted to a complex number. 
It should be stressed that  the sequence of 
limits is first taking $n{\rightarrow}3$ and then
performing $m{\rightarrow}\infty$. If the $1/\epsilon$  pole terms
 arise due  to the UV  divergences, they 
should be first removed  by choosing a subtraction scheme
and by introducing counterterms. Then the 
limit $m{\rightarrow}\infty$ can be performed.

In the background field method we split the gauge field into two 
pieces$\cite{abb}$,
\begin{eqnarray}
{\cal A}_{\mu}=A_{\mu}+Q_{\mu}
\end{eqnarray}
with $A_{\mu}$ being the background field and $Q_{\mu}$ being the quantum one.
Choosing the background covariant gauge condition  $D_{\mu}[A]Q^{\mu}=0$ 
and introducing the 
auxiliary field $B$ to implement Landau type gauge, we obtain the gauge-fixed
effective action,
\begin{eqnarray}
S_{\text{eff}}=S_m[A+Q]+{\int}d^nxB^aD_{\mu}^{ab}[A]Q^{{\mu}b}
+{\int}d^nx\bar{c}^a(D_{\mu}[A]D^{\mu}[A+Q]c)^a,
\end{eqnarray}
where $D_{\mu}^{ab}[A]={\partial}_{\mu}{\delta}^{ab}
+g{\mu}^{\epsilon}f^{acb}A^c_{\mu}$ and $\epsilon=(3-n)/2$.

Feynman rules can be derived in the standard way, which are 
listed as follows:

\begin{itemize}
\item Quantum gauge field propagator\\
\input FEYNMAN
\begin{picture}(20000,800)
\drawline\gluon[\E\REG](8000,300)[6]
\put(6000,500){${\mu},a$}
\put(\particlebackx,500){${\nu},b$}
\end{picture}
\begin{eqnarray}
{\Delta}^{ab}_{\mu\nu}(\tilde{p},\hat{p})&=&
{\langle}Q_{\mu}^a(p)Q_{\nu}^b(-p){\rangle}|_{g=0}
=\frac{{\delta}^{ab}m}{(p^2)^2
+m^2{\tilde{p}}^2}\left[-\mbox{sgn}(k)m{\epsilon}_{\mu\nu\rho}\tilde{p}^{\rho}
+p^2g_{\mu\nu}-p_{\mu}p_{\nu}\right.\nonumber\\[2mm]
&+&\left.\frac{m^2}{p^2}
\left(\frac{\tilde{p}^2}{p^2}p_{\mu}p_{\nu}+\tilde{p}^2\hat{g}_{\mu\nu}
-\hat{p}_{\mu}\hat{p}_{\nu}+p_{\mu}\hat{p}_{\nu}
+\hat{p}_{\mu}p_{\nu}\right)\right];
\label{eq5}
\end{eqnarray}
\item Ghost field propagator\\
\begin{picture}(15000,300)
\drawline\scalar[\E\REG](7000,150)[4]
\put(6000,350){$a$}
\put(\particlebackx,350){$b$}
\end{picture}
\begin{eqnarray}
G^{ab}(p)={\langle}c^a(p)\bar{c}^b(-p){\rangle}|_{g=0}
=\frac{1}{p^2}~{\delta}^{ab};
\end{eqnarray}
\item $QQQ$ and $AQQ$ vertex\\
\begin{picture}(25000,8000)
\drawvertex\gluon[\S 3](6000,5000)[2]
\put(6200,\vertexoney){${\mu},a$}
\put(9000,\vertextwoy){${\rho},c$}
\put(900,\vertexthreey){${\nu},b$}
\drawvertex\gluon[\S 3](20000,5000)[2]
\put(20200,\vertexoney){$\bigcirc\hspace{-4mm}A,{\mu},a$}
\put(23000,\vertextwoy){${\rho},c$}
\put(15000,\vertexthreey){${\nu},b$}
\end{picture}
\begin{eqnarray}
&&-igf^{abc}\left\{\mbox{sgn}(k){\epsilon}_{\mu\nu\rho}
+\frac{{\mu}^{\epsilon}}{m}\left[(p-q)_{\rho}{\delta}_{\mu\nu}
+(q-r)_{\mu}{\delta}_{\nu\rho}+(r-p)_{\nu}{\delta}_{\rho\mu}\right]\right\}
\nonumber\\[2mm]
&&\times (2\pi)^n{\delta}^{(n)}(p+q+r);
\end{eqnarray}
\item $QQQQ$ and $AQQQ$ vertex\\
\begin{picture}(40000,8000)
\drawvertex\gluon[\S 4](10000,5000)[3]
\put(\vertexonex,\vertexoney){${\mu},a$}
\put(\vertextwox,\vertextwoy){${\rho},d$}
\put(\vertexthreex,\vertexthreey){${\lambda},c$}
\put(5500,\vertexfoury){${\nu},b$}
\drawvertex\gluon[\S 4](25000,5000)[3]
\put(\vertexonex,\vertexoney){$\bigcirc\hspace{-4mm}A,\mu,a$}
\put(\vertextwox,\vertextwoy){${\rho},d$}
\put(\vertexthreex,\vertexthreey){${\lambda},c$}
\put(20500,\vertexfoury){${\nu},b$}
\end{picture}
\begin{eqnarray}
&&-\frac{g^2{\mu}^{2\epsilon}}{M}\left[f^{abe}f^{cde}({\delta}_{\nu\rho}
{\delta}_{\mu\lambda}-{\delta}_{\mu\rho}{\delta}_{\nu\lambda})
+f^{ace}f^{bde}({\delta}_{\mu\nu}{\delta}_{\lambda\rho}
-{\delta}_{\mu\rho}{\delta}_{\nu\lambda})\right.\nonumber \\[2mm]
&&+\left.f^{dae}f^{bce}({\delta}_{\nu\rho}{\delta}_{\mu\lambda}
-{\delta}_{\mu\nu}{\delta}_{\lambda\rho})\right]
(2\pi)^n{\delta}^{(n)}(p+q+r+k);
\end{eqnarray}
\item $Ac\bar{c}$ vertex\\
\begin{picture}(10000,8000)
\drawline\gluon[\S\REG](5000,5000)[2]
\put(\gluonfrontx,\gluonfronty){$\bigcirc\hspace{-4mm}A,{\mu},a$}
\drawline\scalar[\SW\REG](\particlebackx,\particlebacky)[2]
\drawarrow[\SW\ATBASE](\pmidx,\pmidy)
\put(1800,\scalarbacky){$b$}
\drawline\scalar[\SE\REG](\gluonbackx,\gluonbacky)[2]
\put(\scalarbackx,\scalarbacky){$c$}
\drawarrow[\SE\ATBASE](\pmidx,\pmidy)
\end{picture}
\begin{eqnarray}
-ig{\mu}^{\epsilon}f^{abc}(p-q)_{\mu}(2\pi)^n{\delta}^{(n)}(p+q+r);
\end{eqnarray}
\item $Qc\bar{c}$ vertex\\
\begin{picture}(10000,8000)
\drawline\gluon[\S\REG](5000,5000)[2]
\put(\gluonfrontx,\gluonfronty){${\mu},a$}
\put(3700,2000){$\cdot$}
\drawline\scalar[\SW\REG](\particlebackx,\particlebacky)[2]
\drawarrow[\SW\ATBASE](\pmidx,\pmidy)
\put(1800,\scalarbacky){$b$}
\drawline\scalar[\SE\REG](\gluonbackx,\gluonbacky)[2]
\drawarrow[\SE\ATBASE](\pmidx,\pmidy)
\put(\scalarbackx,\scalarbacky){$c$}
\end{picture}
\begin{eqnarray}
-ig{\mu}^{\epsilon}f^{abc}p_{\mu}(2\pi)^n{\delta}^{(n)}(p+q+r);
\end{eqnarray}
\item $AAc\bar{c}$ vertex\\
\begin{picture}(20000,8000)
\drawline\gluon[\SE\REG](5000,5000)[2]
\put(\gluonfrontx,\gluonfronty){$\bigcirc\hspace{-4mm}A,{\mu},a$}
\drawline\scalar[\SW\REG](\particlebackx,\particlebacky)[2]
\put(4600,\scalarbacky){$c$}
\drawline\gluon[\NE\REG](\gluonbackx,\gluonbacky)[2]
\put(\gluonbackx,\gluonbacky){$\bigcirc\hspace{-4mm}A,{\nu},b$}
\drawline\scalar[\SE\REG](\gluonfrontx,\gluonfronty)[2]
\put(\scalarbackx,\scalarbacky){$d$}
\end{picture}
\begin{eqnarray}
g^2{\mu}^{2\epsilon}{\delta}_{\mu\nu}(f^{eac}f^{ebd}+f^{ebc}f^{ead})
(2\pi)^n{\delta}^{(n)}(p+q+r+k);
\end{eqnarray}
\item $AQc\bar{c}$ vertex\\
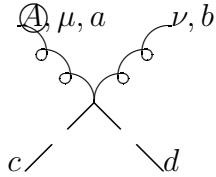
\begin{picture}(20000,8000)
\drawline\gluon[\SE\REG](5000,5000)[2]
\put(\gluonfrontx,\gluonfronty){$\bigcirc\hspace{-4mm}A,{\mu},a$}
\drawline\scalar[\SW\REG](\particlebackx,\particlebacky)[2]
\put(4600,\scalarbacky){$c$}
\drawline\gluon[\NE\REG](\gluonbackx,\gluonbacky)[2]
\put(\gluonbackx,\gluonbacky){${\nu},b$}
\drawline\scalar[\SE\REG](\gluonfrontx,\gluonfronty)[2]
\put(\scalarbackx,\scalarbacky){$d$}
\end{picture}
\begin{eqnarray}
g^2{\mu}^{2\epsilon}{\delta}_{\mu\nu}f^{eac}f^{ebd}
(2\pi)^n{\delta}^{(n)}(p+q+r+k).
\end{eqnarray}
\end{itemize}

Eq.(\ref{eq5}) shows that the quantum gauge field propagator 
at the regularization level takes a very complicated
form due to the dimensional continuation (\ref{eq2}). However, 
it can be decomposed into two parts
\begin{eqnarray}
{\Delta}^{ab}_{\mu\nu}(\tilde{p},\hat{p})=D^{ab}_{\mu\nu}(p)
+R^{ab}_{\mu\nu}(\tilde{p},\hat{p}),
\end{eqnarray}
where
\begin{eqnarray}
D^{ab}_{\mu\nu}(p)&=&\frac{{\delta}^{ab}m}{p^2(p^2+m^2)}
\left[-sgn(k)m{\epsilon}_{\mu\nu\rho}\tilde{p}^{\rho}
+p^2g_{\mu\nu}-p_{\mu}p_{\nu})\right],\nonumber \\[2mm]
R^{ab}_{\mu\nu}(\tilde{p},\hat{p})&=&\frac{{\delta}^{ab}m}
{p^2[(p^2)^2+m^2{\tilde{p}}^2]}\left[\frac{\hat{p}^2}{p^2+m^2}
(-sgn(k)m{\epsilon}_{\mu\nu\rho}\tilde{p}^{\rho}+p^2g_{\mu\nu}
+\frac{m^2}{p^2}p_{\mu}p_{\nu})\right.\nonumber\\[2mm]
&+&\left.{\tilde{p}}^2\hat{\delta}_{\mu\nu}+\hat{p}_{\mu}\hat{p}_{\nu}
-p_{\mu}\hat{p}_{\nu}-\hat{p}_{\mu}p_{\nu}\right].
\end{eqnarray}
A detailed analysis in Ref.\cite{mar1} shows that 
$D^{ab}_{\mu\nu}(p)$ can be 
replaced ${\Delta}^{ab}_{\mu\nu}(\tilde{p},\hat{p})$
as an effective propagator, since $R^{ab}_{\mu\nu}(\tilde{p},\hat{p})$ is 
an evanescent quantity, which has good UV behavior and thus its 
contributions to the loop amplitudes vanish 
in the limit $n{\rightarrow}3$.

\begin{flushleft}
{\bf 3. ~Two-loop Perturbative Analysis and Explicit Calculation}
\end{flushleft}

 The calculation of two-loop Feynman diagrams containing
the propagators and vertices of gauge field is a very heavy work. 
However, since our aim is to verify the perturbative 
finiteness, so our attention is only 
paid on the UV divergent terms. Our strategy is first to show that
the cancellation of the $1/\epsilon$ pole terms in TMYM theory indeed occurs
and then to prove  the existence of the large topological mass, and thus
the two-loop finiteness of CS can be established.

\centerline{\bf A. Cancellation of $1/\epsilon$ Pole Terms in TMYM}

There are several techniques to find out the pole terms:

First, at one-loop level TMYM is finite, 
despite that each Feynman diagram seems
to be divergent from the superficial divergent degree,
the one-loop UV  divergences are actually cancelled. 
Thus there are no need for us to consider the subtraction of 
subdivergence. This
differs from the case of four-dimensional field theory, 
where if the subdivergence cannot 
be well handled, the pole terms with non-polynomial coefficients
will be produced and the unitarity of the theory will be violated. From 
Weinberg's theorem$\cite{iz}$, 
we can see that two-loop divergent behaviour of  TMYM
 is similar to the one-loop behaviour of four-dimensional gauge theory. 
The only problem is whether one can compute the two-loop integral 
analytically.

Second, from the Feynman rules listed in Sect.2, 
one can see that the UV degree of the term involving the antisymmetric 
tensor in quantum effective propagator
${\epsilon}_{\mu\nu\rho}$ is one unit less than that of the whole 
$D^{ab}_{\mu\nu}(p)$ (i.e. -1 and -2, respectively), 
so if ${\omega}$ is the overall
UV degree of a diagram at $n=3$, any integral with $N$ antisymmetric
objects in the diagram amplitude will have an overall UV degree
${\omega}-N$. It is easy to see that the actual 
superficial UV degree ${\omega}$ is zero for every
two-loop diagrams considered in the following. This fact implies that
the integrals exhibiting singularities at $n=3$ only come from pure Yang-Mills
part, i.e. those obtained by formally setting every 
${\epsilon}_{\mu\nu\rho}$ to zero in the amplitude. Furthermore, 
the integral from Yang-Mills part with at least one external momentum  
in the numerator has a negative
UV degree. Therefore, the only sources of UV singularities at $n=3$ are
two-loop integrals which contain neither ${\epsilon}_{\mu\nu\rho}$ nor external
momentum in their numerators. The one-loop finiteness means that there 
exists no subdivergence at $n=3$. Hence the UV singularity is a simple pole at  
$n=3$ independent of $m$ and the external momentum $p$. In addition, the 
dimensional analysis shows that the mass dimension of background field 
two-point function in momentum space is one and so the pole terms must take
the following general form$\cite{mar1}$,
\begin{eqnarray}
A\frac{m}{3-n}g_{\mu\nu},
\end{eqnarray}
where $A$ is a constant to be determined by explicit calculation.

Now let us come to the explicit calculation. 
All the two-loop diagrams contributing to
background two-point function are listed in Fig.2. We shall collect the 
divergent integrals from all of the two-loop diagrams 
and see whether they are cancelled.

We consider two diagrams as examples to illustrate the 
explicit calculation. The amplitude of Fig.2a is read as follows: 
\begin{eqnarray}
 A_{a}&=&g^4\mu^{2\epsilon}
 C_V^2{\delta}^{ab}\left\{ {\int}\frac{d^nq}{(2\pi)^n}
\frac{d^nk}{(2\pi)^n}\frac{4m[k^2q^2-(k{\cdot}q)^2]
(p_{\mu}+2q_{\mu})(p_{\nu}+2q_{\nu})}
{q^4(q+k)^2(p+q)^2k^2(k^2+m^2)}\right.\nonumber\\[2mm]
&+&\left.  {\int}\frac{d^nq}{(2\pi)^n}
\frac{d^nk}{(2\pi)^n}\frac{4m(k^2(p+q){\cdot}(p+q)-[k{\cdot}(p+q)]^2)
(p_{\mu}+2q_{\mu})(p_{\nu}+2q_{\nu}) }
{q^2(p+q+k)^2(p+q)^4k^2(k^2+m^2)}\right\}.
\end{eqnarray}
From the above analysis, we know that the divergent terms are independent of
external momentum and thus to concentrate on the pole terms, we take external 
momentum $p_{\mu}=0$. Rescaling the integration variables $k{\rightarrow}km,
q{\rightarrow}qm$, we obtain
\begin{eqnarray}
 A_a^{\text{(div)}}&=&g^4\mu^{2\epsilon}
C_V^2{\delta}^{ab}{\int}\frac{d^nq}{(2\pi)^n}
\frac{d^nk}{(2\pi)^n}\frac{4m[k^2q^2-(k.q)^2]}{q^6(q+k)^2k^2(k^2+m^2)}q_{\mu}
q_{\nu}\nonumber\\[2mm]
&=& g^4\mu^{2\epsilon} C_V^2{\delta}^{ab}32m^{2n-5}{\int}\frac{d^nq}{(2\pi)^n}
\frac{d^nk}{(2\pi)^n}
\left\{\frac{q_{\mu}q_{\nu}[k^2q^2-(k.q)^2]}{q^6(q+k)^2k^2}-
\frac{q_{\mu}q_{\nu}[k^2q^2-(k.q)^2]}{q^6(q+k)^2(k^2+1)}\right\}
\nonumber\\[2mm]
&=&g^4\mu^{2\epsilon} C_V^2{\delta}^{ab}32m^{2n-5}{\int}\frac{d^nq}{(2\pi)^n}
\frac{d^nk}{(2\pi)^n}\left[-\frac{(k.q)^2}{q^6(q+k)^2k^2}
+\frac{1}{q^4(q+k)^2(k^2+1)}\right.\nonumber\\[2mm]
&+&\left. \frac{(k.q)^2}{q^6(q+k)^2(k^2+1)}\right]
q_{\mu}q_{\nu}.
\label{havethat}
\end{eqnarray}
Using the formula collected in Appendix  and the 
 Feynman parameter integral,
\begin{eqnarray}
\frac{1}{A^mB^n}&=&
\frac{\Gamma(m+n)}{\Gamma(m)\Gamma(n)}{\int}_0^1
dx\frac{x^{m-1}(1-x)^{n-1}}{[Ax+B(1-x)]^{m+n}},\nonumber\\[2mm]
\lim_{{\epsilon}{\rightarrow}0}\Gamma[-{\epsilon}]&=&
-\frac{1}{\epsilon}-\gamma, 
\end{eqnarray}
where $\gamma$ is the Euler constant, it is easy to calculate and see 
that the first term in (\ref{havethat}) 
vanishes, while the second term is 
\begin{eqnarray}
\displaystyle -g^4C_V^2{\delta}^{ab}m\frac{1}{6{\pi}^2}(\frac{1}{\epsilon}
+\gamma){\delta}_{\mu\nu}
\end{eqnarray}
and the third term is
\begin{eqnarray}
g^4C_V^2{\delta}^{ab}m\frac{1}{6{\pi}^2}(-\frac{1}{2}
+\frac{1}{\epsilon}+\gamma){\delta}_{\mu\nu}.
\end{eqnarray}
Thus we have 
\begin{eqnarray}
A_{a}^{\text{(div)}}=0.
\end{eqnarray}

The second example is the amplitude  of Fig.2f,
\begin{eqnarray}
A_f&=&g^4C_V^2{\delta}^{ab}\left\{
{\int}\frac{d^nq}{(2\pi)^n}
\frac{d^nk}{(2\pi)^n}\left[\frac{m(p+2k)_{\nu}
(p+q+k)_{\rho}(k^2{\delta}_{\rho\mu}
-k_{\rho}k_{\mu}) }{q^2(p+q)^2(p+q+k)^2k^2(k^2+m^2)}
\right]\right.\nonumber\\[2mm]
&+&\frac{1}{2}{\int}\frac{d^nq}{(2\pi)^n}
\frac{d^nk}{(2\pi)^n}\left[\frac{mq_{\rho}(k^2{\delta}_{\rho\mu}
-k_{\rho}k_{\mu})(p+2k)_{\nu} }
{(q+k)^2k^2(k^2+m^2)q^2(p+q)^2}\right]\nonumber\\
&+&{\int}\frac{d^nq}{(2\pi)^n}\frac{d^nk}{(2\pi)^n}\left[
\frac{m(2q-p)_{\mu}(2q+k-2p)_{\rho}
(k^2{\delta}_{\nu\rho}-k_{\nu}k_{\rho}) }{q^2(q-p)^2(q-p+k)^2k^2(k^2+m^2)}
\right]\nonumber\\
&+&\left.\frac{1}{2}{\int}\frac{d^nq}{(2\pi)^n}\frac{d^nk}{(2\pi)^n}
\left[\frac{m(2q-p)_{\mu}(k^2{\delta}_{\nu\rho}
-k_{\nu}k_{\rho})q_{\rho}}{(q-p)^2(q+k)^2k^2(k^2+m^2)q^2}\right]\right\}.
\end{eqnarray}
Its divergent part is
\begin{eqnarray}
A_f^{({\rm div})}&=&-5g^4C_V^2{\delta}^{ab}m^{2n-5}{\int}
\frac{d^nq}{(2\pi)^n}
\frac{d^nk}{(2\pi)^n}\frac{2k^2q_{\mu}q_{\nu}
-k.q(q_{\mu}k_{\nu}+k_{\mu}q_{\nu})}{q^4(q+k)^2(k^2+1)}\nonumber\\[2mm]
&=&-5g^4C_V^2{\delta}^{ab}m^{2n-5}{\int}
\frac{d^nq}{(2\pi)^n}\frac{d^nk}{(2\pi)^n}\left[
-\frac{2q_{\mu}q_{\nu}}{q^4(q+k)^2(k^2+1)}+
\frac{1}{2}\frac{q_{\mu}k_{\nu}}{q^2(q+k)^2(k^2+1)}\right.\nonumber\\[2mm]
&-&\left.\frac{1}{2}\frac{q_{\mu}k_{\nu}}{q^4(q+k)^2(k^2+1)}+
\frac{1}{2}\frac{k_{\mu}q_{\nu}}{q^2(q+k)^2(k^2+1)}
-\frac{1}{2}\frac{k_{\mu}q_{\nu}}{q^4(q+k)^2(k^2+1)}\right]\nonumber\\[2mm]
&=&-mg^4C_V^2{\delta}^{ab}\frac{5}{6}\frac{1}{(4\pi)^2}(\frac{1}{\epsilon}
+\gamma).
\end{eqnarray}

The contributions from other diagrams are listed in Table \ref{tab}.
We can see from Table \ref{tab} that the divergent pole terms 
of the two-loop background field two-point function cancel exactly.
Further, recalling
the relation between background field wave function renormalization
constant and vertex renormalization constant$\cite{abb}$, 
\begin{eqnarray}
Z_g=Z_A^{-1/2},
\end{eqnarray}
we can conclude that two-loop background filed 
three-point function also has no $1/{\epsilon}$ pole terms and thus 
the limit ${\epsilon}{\longrightarrow}0$ exists.

\centerline{\bf B. ~Existence of Large Topological Mass Limit}

In last subsection we have shown the cancellation of the 
$1/{\epsilon}$ pole terms by the explicit calculation in TMYM.
However, to prove the two-loop finiteness of CS theory, we need
to show the existence of the large topological mass limit, 
i.e. $m{\rightarrow}{\infty}$.
At first sight, it seems that we should calculate explicitly  the
complicated finite terms of TMYM with respect to the regulator
$\epsilon$ and then study the large $m$-limit. However,
based on the IR finiteness of TMYM in Landau gauge, one can see 
that without needing any explicit calculation a simple analysis can show
that the large $m$-limit indeed exists.

First, we use the fact that in Landau gauge the topological mass
$m$ is the only massive parameter at the two-loop level. Since the $1/\epsilon$
pole terms cancel, the artificial massive parameter $\mu$ has disappeared
after taking the $\epsilon{\longrightarrow}0$ limit. 
So the form factors of the two-loop 
 background field two-point function are only the function of $p^2/m^2$, 
that is, the two-loop background field two-point function
must take the following form,
\begin{eqnarray}
\langle A_{\mu}(p) A_{\nu}(-p) \rangle^{(2)}=
{\epsilon}_{\mu\nu\rho}ip_{\rho}F_1\left[\frac{p^2}{m^2}\right]+
(p^2{\delta}_{\mu\nu}-p_{\mu}p_{\nu})F_2\left[\frac{p^2}{m^2}\right],
\label{eq25}
\end{eqnarray}   
where we consider the explicit gauge symmetry of background field 
two-point function.

Second, we make use of a key property of TMYM in Landau gauge$\cite{rao}$:
TMYM is infrared finite to every order of perturbative theory. 
This means that the limit $p^2{\longrightarrow}0$ exists, since
the form factors $F_1$ and $F_2$ of background field two-point function
can only be the functions of $p^2/m^2$, the limit $p^2{\longrightarrow}0$
is equivalent to large-$m$ limit. Of course, one may
think that there could exist the terms with the following form factors:
\begin{eqnarray}
\frac{m^l}{(1+p^2/m^2)^n}, ~~l>0,
\end{eqnarray}  
for which the limit $p^2{\longrightarrow}0$ is 
not equivalent to large-$m$ limit. 
However, since the mass dimension of  
$\langle A_{\mu}(p) A_{\nu}(-p) \rangle $ is one, we can 
see from Eq.(\ref{eq25}) that this is not possible
for $l>0$, and hence the terms with this kind of form factors 
do not exist. Therefore, the
large topological mass limit indeed exists.

\begin{flushleft}
{\bf 4.~Conclusion}
\end{flushleft}

Background field  method is an elegant calculation technique in 
gauge field theories. The
explicit gauge symmetry of background field is preserved if the 
background covariant gauge is chosen. The calculations are then reduced 
considerably. In this paper we have applied this method to the 
investigation of two-loop perturbative finiteness of 
CS theory in the concrete regularization scheme, namely the 
hybrid regularization of higher-covariant derivative and dimensional 
regularization. By explicit calculations we show that the pole terms with
respect to the regulator $\epsilon$ cancel 
exactly. Further, based on the gauge symmetry, the dimensional
analysis and the IR finite property of TMYM, we show that
the limit of large topological mass also exists. Thus
the two-loop finiteness of CS theory has been explicitly proven. This has
not only  verified the reasonableness of non-perturbative
analysis$\cite{blasi}$, but has also provided a strong support
to the description of link polynomial in terms of CS theory
and the Witten conjecture $\cite{witt}$ about the relation between CS 
and two-dimensional conformal field theories.

\noindent{\bf Acknowledgment:}\\
The  financial support of the Academy of Finland under the Project No. 37599
is greatly acknowledged. W.F.C. thanks the World
Laboratory, Switzerland for financial support.
We would like to thank Dr. F. Ruiz Ruiz for enlightening and 
useful discussions.

\appendix
\setcounter{equation}{0}

\section{}

Some formulas used in the two-loop integrations are collected in this appendix:
\begin{eqnarray}
&&\int\frac{d^nq}{(2\pi)^n}\frac{d^nk}{(2\pi)^n}
\frac{q_{\mu}k_{\nu}}{(q^2)^{\alpha}[(q+k)^2]^{\beta}(k^2+1)^{\gamma}}
\nonumber\\[2mm]
&&=-\frac{1}{(4\pi)^n}\frac{{\delta}_{\mu\nu}}{n}
\frac{\Gamma({\alpha}+{\beta})
\Gamma(n/2-{\alpha}+1)\Gamma(n/2-{\beta})\Gamma({\alpha}+{\beta}+{\gamma}-1-n)}{
\Gamma({\alpha})\Gamma({\beta})\Gamma({\gamma})\Gamma(n/2)}.
\end{eqnarray}

\begin{eqnarray}
&&\int\frac{d^nq}{(2\pi)^n}\frac{d^nk}{(2\pi)^n}
\frac{k_{\mu}q_{\nu}}{(q^2)^{\alpha}(k^2)^{\beta}[(q+k)^2+1]^{\gamma}}
\nonumber\\[2mm]
&&=-\frac{1}{(4\pi)^n}\frac{{\delta}_{\mu\nu}}{n}
\frac{\Gamma({\beta}+{\gamma}-n/2)
\Gamma(1+n/2-{\beta})\Gamma(1+n/2-\alpha)
\Gamma({\alpha}+{\beta}+{\gamma}-1-n)}
{\Gamma({\alpha})\Gamma({\beta})\Gamma({\gamma})\Gamma(n/2)}\nonumber\\[2mm]
&&{\times}\Gamma({\alpha}+{\beta}-n/2-1).
\end{eqnarray}

\begin{eqnarray}
&&\int\frac{d^nq}{(2\pi)^n}\frac{d^nk}{(2\pi)^n}
\frac{k_{\mu}q_{\nu}}{(q^2+1)^{\alpha}(k^2)^{\beta}[(q+k)^2+1]^{\gamma}}
\nonumber\\[2mm]
&&= -\frac{1}{(4\pi)^n}\frac{{\delta}_{\mu\nu}}{n}
\frac{\Gamma(\beta+2\gamma-n/2)\Gamma(1+n/2-\beta)
\Gamma(\alpha+\beta+2\gamma-n-1)}
{\Gamma({\alpha})\Gamma(\gamma)\Gamma(n/2)\Gamma(\alpha+2\beta+2\gamma-n-1)}
\nonumber\\[2mm]
&&{\times}\Gamma(\alpha+\beta+2\gamma-n/2-1).
\end{eqnarray}

\begin{eqnarray}
&&\int\frac{d^nq}{(2\pi)^n}\frac{d^nk}{(2\pi)^n}
\frac{k_{\mu}k_{\nu}}{(q^2+1)^{\alpha}(k^2+1)^{\beta}[(q+k)^2]^{\gamma}}
=\frac{1}{(4\pi)^n}\frac{{\delta}_{\mu\nu}}{n}\nonumber\\[2mm]
&&{\times}\left[
\frac{\Gamma(\alpha+\gamma-n/2)\Gamma(n/2-\gamma)\Gamma(\beta+\gamma-n/2-1)
\Gamma(\alpha+\beta+\gamma-n-1)}{\Gamma(\alpha)\Gamma(n/2)\Gamma(\beta-1)
\Gamma(\alpha+2\beta+2\gamma-n-1)}\right.\nonumber\\[2mm]
&&\left.-\frac{\Gamma(\alpha+\gamma-n/2)\Gamma(n/2-\gamma)
\Gamma(\beta+\gamma-n/2)
\Gamma(\alpha+2\beta+2\gamma-n)}{\Gamma(\alpha)\Gamma(n/2)\Gamma(\beta)
\Gamma(\alpha+\beta+2\gamma-n)}\right].
\end{eqnarray}

\begin{eqnarray}
&& \int\frac{d^nq}{(2\pi)^n}\frac{d^nk}{(2\pi)^n}
\frac{q_{\mu}q_{\nu}}{(q^2)^{\alpha}[(q+k)^2]^{\beta}(k^2+1)^{\gamma}}=
\frac{1}{(4\pi)^n}\frac{{\delta}_{\mu\nu}}{n}\frac{\Gamma(n-\alpha-\beta+1)
\Gamma(\alpha+\beta+\gamma-n-1)}{[\Gamma(n/2)]^2\Gamma(\alpha)
\Gamma(\beta)\Gamma(\gamma)}\nonumber\\[2mm]
&&{\times}\left[\frac{\Gamma(n/2+1)\Gamma(\alpha+\beta-n/2-1)
\Gamma(n/2-\beta+1)\Gamma(n/2-\alpha+1)}{\Gamma(n-\alpha-\beta+2)}\right.\\[2mm]
&&\left.+
\frac{\Gamma(n/2)\Gamma(\alpha+\beta-n/2)\Gamma(n/2-\alpha+2)\Gamma(n/2-\beta)}
{\Gamma(n-\alpha-\beta+2)}\right].
\end{eqnarray}

\begin{eqnarray}
&&\lim_{n{\rightarrow}3}\int\frac{d^nq}{(2\pi)^n}
\frac{d^nk}{(2\pi)^n}\frac{k_{\mu}q_{\nu}}{(q^2+1)(k^2+1)[(q+k)^2+1]}
\nonumber\\[2mm]
&&=\lim_{n{\rightarrow}3}\int\frac{d^nq}{(2\pi)^n}
\frac{d^nk}{(2\pi)^n}\frac{q_{\mu}k_{\nu}}{(q^2+1)(k^2+1)[(q+k)^2+1]}
\nonumber\\[2mm]
&&=-\frac{{\delta}_{\mu\nu}}{3}\left[\frac{1}{16{\pi}^2}
-\frac{1}{32{\pi}^2}(-\frac{1}{\epsilon}+1
-\gamma+\ln(\frac{4\pi}{9}))\right].
\end{eqnarray}

\begin{eqnarray}
&&\lim_{n{\rightarrow}3}\int\frac{d^nq}{(2\pi)^n}
\frac{d^nk}{(2\pi)^n}\frac{q_{\mu}q_{\nu}}{(q^2+1)(k^2+1)[(q+k)^2+1]}
\nonumber\\[2mm]
&&=\frac{{\delta}_{\mu\nu}}{3}\left[-\frac{1}{16{\pi}^2}+\frac{1}{32{\pi}^2}
(-\frac{1}{\epsilon}+1-\gamma+\ln(\frac{4\pi}{9}))\right].
\end{eqnarray}

\newpage

\begin{table}
\begin{center}

\begin{tabular}{|c|c|c|c|c|c|c|c|c|c|c|}
 $a$ & $b$ & $c$ & $d$ & $e$ & $f$ & $g$ & $h$ & $i$ & $j$ & $k$\\ \hline
 $0$ & $\displaystyle\frac{19}{4}\frac{1}{\epsilon}$ &
  $\displaystyle\frac{1}{3}\frac{1}{\epsilon}$ &
   -$\displaystyle\frac{7}{2}\frac{1}{\epsilon}$ &
   $0$ & -$\displaystyle\frac{5}{6}\frac{1}{\epsilon}$ &
  $0$ & $\displaystyle\frac{8}{3}\frac{1}{\epsilon}$ &
   $\displaystyle\frac{3}{4}\frac{1}{\epsilon}$ &
  $\displaystyle\frac{1}{3}\frac{1}{\epsilon}$ &
  -$\displaystyle\frac{9}{2}\frac{1}{\epsilon}$ \\ \hline
total & \multicolumn{10}{l|}{$a+b+c+d+e+f+g+h+i+j+k=0$}
\end{tabular}

\caption{\protect\small Pole terms from two-loop diagrams (in the units of
$1/(4\pi)^2\,g^4C_V^2$)}
\label{tab}
\end{center}
\end{table}

\begin{figure}

\vspace{20cm}

\caption{\protect\small Two-loop Feynman diagrams}

\end{figure}
 
\end{document}